# A Business Intelligence Model to Predict Bankruptcy using Financial Domain Ontology with Association Rule Mining Algorithm

A.Martin [1], M.Manjula [2] and Dr.Prasanna Venkatesan [3]

[1] Research Scholar, Dept.of Banking Technology,
Pondicherry University,
Puducherry.

[2] PG Student, Dept.of Computer Science and Engg,
Sri Manakula Vinayagar Engg College, Madagadipet,
Puducherry.

[3] Associate Professor, Dept.of Banking Technology,
Pondicherry University,
Puducherry.

**Abstract**
Today in every organization financial analysis provides the basis for understanding and evaluating the results of business operations and delivering how well a business is doing. This means that the organizations can control the operational activities primarily related to corporate finance. One way that doing this is by analysis of bankruptcy prediction. This paper develops an ontological model from financial information of an organization by analyzing the Semantics of the financial statement of a business. One of the best bankruptcy prediction models is Altman Z-score model. Altman Z-score method uses financial rations to predict bankruptcy. From the financial ontological model the relation between financial data is discovered by using data mining algorithm. By combining financial domain ontological model with association rule mining algorithm and Z-score model a new business intelligence model is developed to predict the bankruptcy.

## 1. Introduction

The Bankruptcy Prediction is very important for any organization. For predicting the bankruptcy the financial statement is used. The financial analysis is used for the analysis of the financial statement. In general the financial statement has both balance sheet and income statement. Based on the financial statement, a bankruptcy prediction model is proposed to predict the bankruptcy. Ontology models address a standard to provide more insightful semantics and sharper level of representation for financial analysis .Altman Z-score method is one of the best methods used in predicting the potential bankruptcy of a company that consists of a combination of five financial ratios. In this paper we propose a Data Mining algorithm - Association Rule Mining Algorithm to extract the new knowledge from the relevant data obtained through financial ontology tree, to come up with the necessary decisions.

The rest of this paper is structured as follows. Section 2 describes the Prior Research on Ontology. Section 3 explains the Ontology Model. Section 4 explains the Data Mining Techniques. Section 5, the Bankruptcy Prediction Model .Section 6 describes the Business Intelligence Model. Section 7 discusses our Approach, followed by conclusions in Section 8.

## 2. Prior Research on Ontology

Query interfaces provided by the Deep Web are clues to disclose the hidden schemas. But the complicated semantic relationships in the query interfaces lead to the lower generality and ability of local-as-view (LAV) method in the traditional information fusion system. To address the problem, an Ontology extended semantic related group algorithm is presented. The algorithm is to extract attributes using Ontology technique, and the semantic distance between any pairs of candidate attributes is evaluated by Word Net. A semantic matrix is generated and a reverse backtracking algorithm is used to find semantic related groups of attributes. A matching mechanism is constructed on the semantic related groups, by which the expression ability of LAV is extended. The validation and effectiveness were tested based on experiments [1].

Information integration and interoperability among information sources are related problems that have received significant attention since early days of computer information processing. The objective of Web information integration system is to promote Web information collection, sharing and retrieval in distributed and heterogeneous environment. This paper puts forward an ontology-based approach to Web information integration, which promises better access to relevant information by producing a global domain ontology that is used for establishing semantic connections among the local autonomous information sources as well as for retrieving those sources. In this approach, an up-bottom information integration method with a semantic transformation adapter framework is proposed to solve both the semantic heterogeneities and the query answering problems [2].

## 3. Ontology

Ontology is a formal, explicit specification of a shared conceptualization. Conceptualization refers to an abstract model of some phenomenon in the world by identifying the relevant concepts of that phenomenon. Explicit Specification means that the type of concepts used and the constraints on their use are explicitly defined. Formal refers to the fact that the ontology should be machine-readable. Shared reflects the notion that an ontology captures consensual knowledge that is not private to some individual, but accepted by a group[3].

Ontology comprises of classes of entities, relations between entities and the axioms, which apply to the entities of that domain. It is made up of the following parts:

- *Classes and instances:* describe the sets, collections, concepts, types of objects, or kinds of things, for example, an ontology model financial domain structure may contain classes such as "firm's category".
- *Properties:* They establish relationships between the concepts of ontology.
- *Rules:* They model logical sentences that are always true. Rules provide us with high expressiveness and they make for more complex reasoning with the ontology, which can give rise to scalability issues.

*3.*1 Relations between Ontology Models

There are four kinds of Ontologies are available.The relations between these Ontology model is depicted in figure 1[4].

- ***Top-level ontologies*:** describe very general concepts like *space*, *time*, *event*, which are independent of a particular domain. It seems reasonable to have unified top-level ontologies for large communities of users.
- ***Domain ontologies*:** describe the vocabulary related to a generic domain by specializing the concepts introduced in the top-level ontology.
- ***Task ontologies*:** describe the vocabulary related to a generic task or activity by specializing the top-level ontologies.
- ***Application ontologies*:** describe the most specific ones. Concepts in application ontologies often correspond to roles played by domain entities.

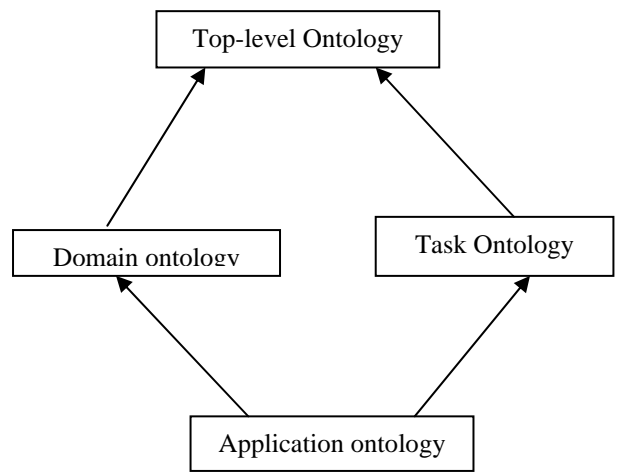

Fig. 1  Relations Between Ontology Models

This paper address about domain Ontology. Using domain Ontology the financial Ontological tree is Constructed.

3.2 Domain Ontologies

Domain ontologies define domain specific conceptualizations and impose descriptions on the domain knowledge structure and content. They refer to the detailed structuring of a context of analysis with respect to the sub-domains, which it is composed of grant sub-domains [3].

An effective ontological domain analysis clarifies the terminology, which enables the ontology to work for coherent and cohesive reasoning purposes. Shared ontologies can form the basis for domain-specific knowledge representation languages.

## 3.2 Example of Domain Ontologies

A financial statement that summarizes a company's assets, liabilities and shareholders' equity at a specific point in time. A balance sheet is a snapshot of a business' financial condition at a specific moment in time, usually at the close of an accounting period. A balance sheet comprises assets, liabilities, and owners' or stockholders' equity. Assets and liabilities are divided into short- and long-term obligations including cash accounts such as checking, money market, or government securities. At any given time, assets must equal liabilities plus owners' equity. An asset is anything the business owns that has monetary value. Liabilities are the claims of creditors against the assets of the business.

The balance sheet must follow the following formula:

$$\text{Assets} = \text{Liabilities} + \text{Shareholders Equity}$$

*Assets*- Assets are subdivided into current and long-term assets to reflect the ease of liquidating each asset. Cash, for obvious reasons, is considered the most liquid of all assets. Long-term assets, such as real estate or machinery, are less likely to sell overnight or have the capability of being quickly converted into a current asset such as cash.

*Current Assets*-Current assets are any assets that can be easily converted into cash within one calendar year. Examples of current assets would be checking or money market accounts, accounts receivable, and notes are

- *Cash* -Money available immediately, such as in checking accounts, is the most liquid of all short-term assets.

- *Accounts receivables* -This is money owed to the business for purchases made by customers, suppliers, and other vendors.

- *Notes receivables*-Notes receivables that are due within one year are current assets. Notes that cannot be collected on within one year should be considered long-term assets.

*Fixed assets* -Fixed assets include land, buildings, machinery, and vehicles that are used in connection with the business.

- *Land* -Land is considered a fixed asset but, unlike other fixed assets, is not depreciated, because land is considered an asset that never wears out.

- *Buildings*- Buildings are categorized as fixed assets and are depreciated over time.

- *Office equipment* -This includes office equipment such as copiers, fax machines, printers, and computers used in your business.

- *Machinery* -This figure represents machines and equipment used in your plant to produce your product. Examples of machinery might include lathes, conveyor belts, or a printing press.

- *Vehicles* -This would include any vehicles used in your business.

- *Total fixed assets* -This is the total dollar value of all fixed assets in your business, less any accumulated depreciation.

*Total assets*- The total dollar value of both the short-term and long-term assets of your business.

*Liabilities and owners' equity* -This includes all debts and obligations owed by the business to outside creditors, vendors, or banks that are payable within one year, plus the owners' equity. Often, this side of the balance sheet is simply referred to as "Liabilities."

- *Accounts payable*- This is comprised of all short-term obligations owed by your business to creditors, suppliers, and other vendors. Accounts payable can include supplies and materials acquired on credit.

- *Notes payable* -This represents money owed on a short-term collection cycle of one year or less. It may include bank notes, mortgage obligations, or vehicle payments.

*Accrued payroll and withholding*- This includes any earned wages or withholdings that are owed to or for employees but have not yet been paid

- *Total current liabilities*- This is the sum total of all current liabilities owed to creditors that must be paid within a one-year time frame.

- *Long-term liabilities*- These are any debts or obligations owed by the business that are due more than one year out from the current date.

- *Mortgage note payable-* This is the balance of a mortgage that extends out beyond the current year. For example, you may have paid off three years of a fifteen-year mortgage note, of which the remaining eleven years, not counting the current year, are considered long-term.

- *Owners' equity-* Sometimes this is referred to as stockholders' equity. Owners' equity is made up of the initial investment in the business as well as any retained earnings that are reinvested in the business.

- Common stock -This is stock issued as part of the initial or later-stage investment in the business.

- *Retained earnings* -These are earnings reinvested in the business after the deduction of any distributions to shareholders, such as dividend payments.

*Total liabilities and owners' equity* -This comprises all debts and monies that are owed to outside creditors, vendors, or banks and the remaining monies that are owed to shareholders, including retained earnings reinvested in the business.

For this system a balance sheet is given as input. Any form of balance sheet or income statement can be given as input to this system. A sample balance sheet is given in table 1.

|  | Feb 2008 | Feb 2009 | Dec 2009 | Jan 2010 | Feb 2010 | PriorMo | YTD | PriorYr |
|---|---|---|---|---|---|---|---|---|
| ASSETS | | | | | | | | |
| Current Assets | | | | | | | | |
| Cash in Bank s | 60,650 | 97,957 | 132,251 | 123,558 | 130,490 | 5.6% | -1.3% | 33.2% |
| Accounts Receivable | 10,846 | 32,778 | 6,775 | 11,074 | 12,663 | 14.3% | 86.9% | -61.4% |
| Total Current Assets | 71,496 | 130,735 | 139,026 | 134,633 | 143,153 | 6.3% | 3.0% | 9.5% |
| Long-term Assets | | | | | | | | |
| Investments | 6,157 | 7,814 | 37,583 | 39,211 | 40,707 | 3.8% | 8.3% | 420.9% |
| Retirement | 65,055 | 54,829 | 127,279 | 123,912 | 128,891 | 4.0% | 1.3% | 135.1% |
| 2004 Honda Civic | 8,778 | 7,773 | 6,898 | 6,898 | 6,898 | 0.0% | 0.0% | -11.3% |
| Total Long-term Assets | 79,989 | | 171,760 | 170,021 | 176,496 | 3.8% | 2.8% | 150.6% |
| Total Assets | 151,486 | 201,151 | 310,786 | 304,654 | 319,649 | 4.9% | 2.9% | 58.9% |
| LIABILITIES | | | | | | | | |
| Current Liabilities | | | | | | | | |
| Accounts payable | 5,753 | 1,590 | 10,027 | 6,541 | 5,125 | 21.7% | 48.9% | -223.3% |
| Total Current Liabilities | 5,753 | 1,590 | 10,027 | 6,541 | 5,125 | 21.7% | 48.9% | -223.3% |
| Long-term Liabilities | | | | | | | | |
| Student Loans | 12,814 | 0 | 0 | 0 | 0 | | | |
| Total long-term Liabilities | 12,814 | 0 | 0 | 0 | 0 | | | |
| TOTAL LIABILITIES | 18,567 | 1,590 | 10,027 | 6,541 | 5,125 | 21.7% | 48.9% | -233.3% |
| OVERALL TOTAL | 132,919 | 199,561 | 300,759 | 298,113 | 314,525 | 5.5% | 4.6% | 57.6% |

*Table 1: Balance sheet*

The Financial Domain Ontology is constructed for a given financial statement (i.e.) which may be a balance sheet or income statement. The Example of the *balance sheet* ontology might be depicted in Protégé tool as shown in figure 2 [5].

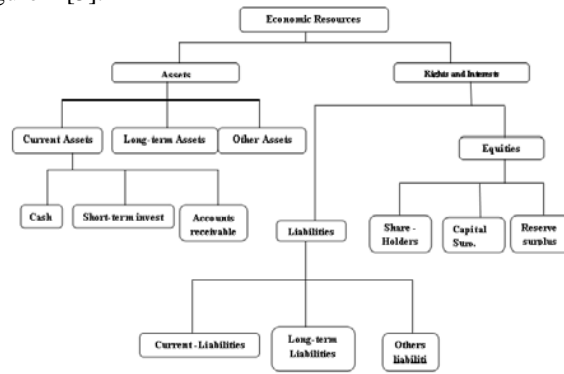

Figure 2: Financial Ontological tree

Financial ontology is made of Classes, Instances, Slots and Facets. Classes represent the knowledge concept in financial domain, the instances are the instances of concept, slots describe the relations between concepts and may be referenced by classes and instances, facets is the constraints of slots and is applied to define the functions of slots [6]. For example, every financial ratio is a class, and includes two attributes (slots) which stand for two accounting items in financial statement.

The following XML coding document is a snippet of *Finance* ontology represented by OWL [4].

```xml
<?xml version="1.0"?>
<rdf:RDF
xmlns:protege="http://protege.stanford.edu/plugins/owl/prote
<owl:Ontology rdf:about="">
<owl:imports
rdf:resource="http://protege.stanford.edu/plugins/owl/protege"/>
</owl:Ontology>
<owl:Class rdf:ID="Asset">
<rdfs:comment xml:lang="en"> valuable items that is
</rdfs:comment>
</owl:Class>
<owl:Class rdf:ID="Loan">
<rdfs:subClassOf>
<owl:Class rdf:about="#MortgageLoan"/>
</rdfs:subClassOf>
</owl:Class>
<owl:Class rdf:ID="ProductRateApplicationVariable">
<owl:disjointWith>
<owl:Class rdf:about="#ProductRateApplicationFixed"/>
</owl:disjointWith>
<rdfs:subClassOf>
<owl:Class rdf:about="#ProductRateApplication"/>
</rdfs:subClassOf>
</owl:Class>
<owl:Class rdf:ID="Customer">
<rdfs:comment xml:lang="en">enterprise client, who usuall
contractual relationship with the enterprise</rdfs:comment>
<rdfs:subClassOf rdf:resource="#User"/>
</rdf:RDF>
```

## 4. Data Mining

Data mining, the extraction of the hidden predictive information from large databases, is a powerful new technology with great potential to hidden Knowledge in the data warehouse. Here Data Mining Algorithm is applied to ontological tree to mine the hidden relations. In this paper to discover the new knowledge data mining is applied in two steps. First cluster analysis is performed to group the given balance sheet into respective cluster. Once cluster is developed association rule mining algorithm is applied to discover the new knowledge for decision making.

### 4.1 Data Mining Techniques

*Clustering analysis:* Clustering Analysis segments a large set of data into subsets or clusters. Each cluster is a collection of data objects that are similar to one another within the same cluster but dissimilar to objects in other clusters. In other words, objects are clustered based on the principle of maximizing the intra-class similarity while minimizing the inter-class similarity. For example, clustering techniques can be used to identify stable dependencies for risk management and investment management.

*Association rule mining:* Association rule mining uncovers interesting correlation patterns among a large set of data items by showing attribute- value conditions that occur together frequently. A typical example is market basket analysis, which analyzes purchasing habits of customers by finding associations between different items in customers' "shopping baskets."

There are several algorithms are available for supporting Association rules, some of the algorithms are:
1. A priori algorithm
2. Partition algorithm
3. Pincer – search algorithm
4. FP- Tree Growth algorithm

*1. A priori Algorithm*
A priori is the best-known algorithm to mine association rules. It uses a breadth-first search strategy to counting the support of item sets and uses a candidate generation function which exploits the downward closure property of support.

*2. Partition algorithm*
The partition algorithm is based on the observation that the frequent sets are normally very few number compared to the sets of all item sets.

*3. Pincer – search algorithm*
Pincer Search combines the following two approaches:
*Bottom-up:* Generate subsets and go on to generating parent-set candidate sets using    frequent subsets.
*Top-Down:* Generating parent set and then generating subsets.
 It also uses two special properties:
 *Downward Closure Property:* If an itemset is frequent, then all its must be frequent.
 *Upward Closure Property:* If an itemset is infrequent, all its supersets must be infrequent.
*4. FP- Tree Growth algorithm*
FP-growth (frequent pattern growth) uses an extended prefix-tree (FP-tree) structure to store the database in a compressed form. FP-growth adopts a divide-and-conquer approach to decompose both the mining tasks and the databases.

In this project, the A priori Algorithm is applied because the A priori Algorithm is the best algorithm for finding Frequent itemsets in the database. Frequent itemsets are generated by finding the relationships among the data.

*A Priori Algorithm:*

An association rule mining algorithm (A priori) has been developed for rule mining in large transaction databases [7]. An *item set* is a non-empty set of items.

They have decomposed the problem of mining association rules into two parts.

- Find all combinations of items that have transaction support above minimum support. Call those combinations frequent item sets.
- Use the frequent item sets to generate the desired rules. The general idea is that if, say, ABCD and AB are frequent item sets, then we can determine if the rule AB CD holds by computing the ratio r = support(ABCD)/support(AB). The rule holds only if r >= minimum confidence. Note that the rule will have minimum support because ABCD is frequent. The algorithm is highly scalable [8].

A priori algorithm used in Quest for finding all frequent item sets is given below.

**procedure** A prioriAlg()

**begin**
L1 := {frequent 1-itemsets};
**for** ( k := 2; Lk-1 0; k++ ) **do** {
Ck= apriori-gen(Lk-1) ; // new candidates
**for** all transactions t in the dataset **do** {
**for** all candidates c Ck contained in t **do**
c:count++
}
Lk = { c Ck | c:count >= min-support}
}
Answer: = k Lk
**end**

It makes multiple passes over the database. In the first pass, the algorithm simply counts item occurrences to determine the frequent 1-itemsets (item sets with 1 item). A subsequent pass, say pass k, consists of two phases.

First, the frequent item sets Lk-1 (the set of all frequent (k-1)-item sets) found in the (k-1)th pass are used to generate the candidate item sets Ck, using the a priori-gen() function. This function first joins Lk-1 with Lk-1, the joining condition being that the lexicographically ordered first k-2 items are the same. Next, it deletes all those item sets from the join result that have some (k-1)-subset that is not in Lk-1 yielding Ck. The algorithm now scans the database. For each transaction, it determines which of the candidates in Ck are contained in the transaction using a hash-tree data structure and increments the count of those candidates. At the end of the pass, Ck is examined to determine which of the candidates are frequent, yielding Lk. The algorithm terminates when Lk becomes empty [7].

## 5. Bankruptcy

A person or a business unable to repay its outstanding debts [9] is called bankruptcy. To assess the economic health of the company bankruptcy is very much useful.

5.1 Bankruptcy Prediction Model

Bankruptcy predicting models, derived from these financial statement ratios, assist shareholders, stakeholders, company managers, and other directly and indirectly related entities such as suppliers, customers, and competitors in predicting financial problems of a company. This helps the companies to plan their strategies and to know the strengths and weakness of related companies and act accordingly. This is crucial for the company success. However, there are three main problems that old bankruptcy predication models may not be accurate predictors on services and information technology companies [9].

- First, the bankruptcy prediction models such as Altman were developed when manufacturing companies were dominant in the market, which is not true at present.
- Second, the service and information technology companies are characterized by a different set of financial norms than the manufacturing companies.
- The third problem is the effect of rapid changes in the services and information technology companies that makes bankruptcy prediction more difficult and complicated.

Therefore, there is a need to investigate whether these Altman models are still applicable in order to assist financial institutions, banks, and other organizations to predict failure accurately in the service and information technology companies.

## 5.2 Altman's Z-score Model

Edward Altman has developed a model using financial statement ratios and multiple discriminate analyses to predict bankruptcy for publicly traded manufacturing firms. The resultant model is of the form [10].

The final discriminate function is as follows:

$Z = 0.012X1 + 0.014X2 + 0.033X3 + 0.006X4 + 0.999X5$
Where
 X1 = working capital/total assets,
X2 = retained earnings/total assets,
X3 = earnings before interest and taxes/total assets,
X4 = market value equity/book value of total liabilities,
X5 = sales/total assets, and
Z = overall index.

A score of Z less than 2.675 indicates that a firm has a 95 percent chance of becoming bankrupt within one year. However, Altman's results show that in practice scores between 1.81 and 2.99 should be thought of as a gray area. In actual use, bankruptcy would be predicted. if Z < = 1.81 and non-bankruptcy  if  Z >=2.99. Altman shows that bankrupt firms and non-bankrupt firms have very different financial profiles one year before bankruptcy [11].

The resultant Business Intelligence model integrates ontology model, data mining model to find the knowledge of success and failure of an organization and classifying the organization as bankrupt or not bankrupt. The architecture of business intelligence model is depicted in the following section.

## 6. Architecture of BI Model for Bankruptcy Prediction

Business Intelligence (BI) is a terminology representing a collection of processes, tools and technologies helpful in achieving more profit by considerably improving the productivity, sales and service of an enterprise. With the help of BI methods, the corporate data can be organized, analyzed in a better way and then converted into a useful knowledge of information needed to initiate a profitable business action. Thus it's about turning a raw, collected data into intelligent information by analyzing and re-arranging the data according to the relationships between the data items by knowing what data to collect and manage and in what context.

The financial data from various business financial statements (i.e.) Balance sheet, income statement are collected.   Based on the financial data the financial domain ontology tree is generated. By combining the Data Mining algorithm and Bankruptcy Prediction model the Business intelligence (BI) is generated. The BI is used to report whether or not the business is going to be bankrupt.

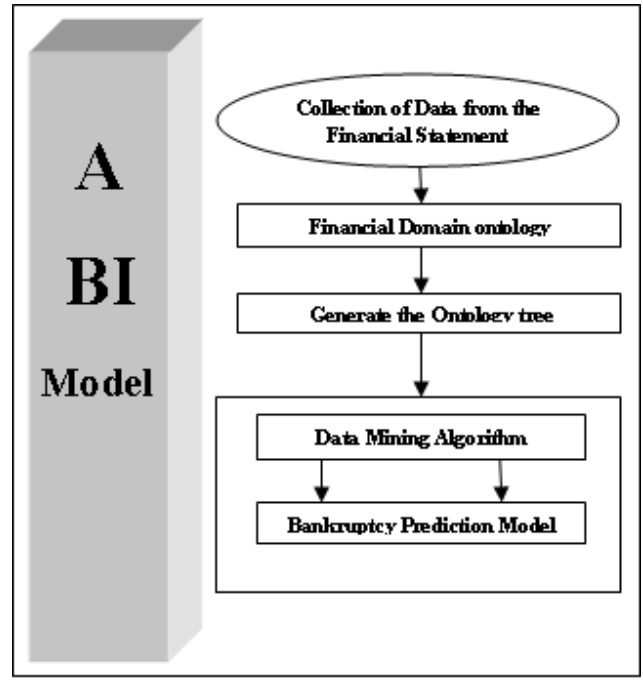

Figure 3: A BI model for Bankruptcy prediction

Companies now have massive databases, which are growing, but only a limited amount of the data is used in the decision-making process. Businesses require data systems that can handle the large volume of data and transform it into valuable information that can be used by business users. There are various structures that are applied in the architecture of business intelligence. These include operational data store, data warehouse and data mart.

With the operational data store, users get a consistent view of the data that is currently stored in the database. This structure is controlled using a transcription processing system. When data is changed in the source systems, you get a copy of the data that has been changed in the operational data store. The data that exists in the operational data store changes to reflect the data that is currently found in your source system. In most cases, the storage process is in real time and can be applied in the various operations carried out to run a business.

## 7. Conclusion

The use of ontologically clustered data mining Bankruptcy prediction BI Model has an advantage over the existing Bi models since the semantics based decisions yield better results to syntactic based models. The Association Rule mining Algorithm augments the efficiency of the proposed method by providing relevant results based on the association between the businesses' financial statements. By integrating ontological model with data mining Algorithm and Bankruptcy prediction model a new business intelligence model can be used to effectively.

**First Author** Mr.A.Martin Assistant Professor in the Department of Information Technology inSri Manakula Vinayagar Engineering College. Pondicherry University,Pudhucherry, India. He holds a M.E and pursuing his Ph.D in Banking Technology from Pondicherry University, India.

**Second Author** Ms.M.Manjula II year M.tech student in the project phase in the Department of computer Science and Engineering in Sri Manakula Vinayagar Engineering College. Pondicherry University, Pudhucherry.

**Third Author** Dr.V.Prasanna Venkatesan, Associate Professor, Dept. of Banking Technology, Pondicherry University, Pondicherry. He has more than 20 years Teaching and Research experience in the field of Computer Science and Engineering; He has developed a compiler for multilingual languages.